# High Speed All-optical extended DV-Curve-based DNA sequence alignment utilizing wavelength and polarization modulation

EHSAN MALEKI[1], HOSSEIN BABASHAH[2], SOMAYYEH KOOHI[1,*], ZAHRA KAVEHVASH[2]

[1]*Department of Computer Engineering, Sharif University of Technology, Tehran, Iran*
[2]*Department of Electrical Engineering, Sharif University of Technology, Tehran, Iran*
*\*Corresponding author: koohi@sharif.edu*

This paper presents a novel optical processing approach for exploring genome sequences built upon optical correlator for global alignment and extended DV-curve method for local alignment. To overcome the problem of traditional DV-curve method for presenting an accurate and simplified output, we propose HAWPOD, built upon DV-curve method, to analyze genome sequences in five steps: DNA coding, alignment, noise cancellation, simplification, and modification. Moreover, all-optical implementation of the HAWPOD method is developed, while its accuracy is validated through numerical simulations in LUMERICAL FDTD. The results express the proposed method is much faster than its electrical counterparts, such as Basic Local Alignment Search Tools.

## 1. INTRODUCTION

Bioinformatics is an interdisciplinary field that expands methods and tools for biological data comprehension, and benefits from several sciences, such as computer science, statistics, mathematics, and engineering to expound and analyze biological data [1]. As an important operation in bioinformatics, sequence alignment arranges the sequences of DNA to identify regions of similarity and differences between the sequences [2] to detect the genetic disease [3]. Each DNA sequence is represented with a string of A/G/T/C characters; each is called a nucleotide, and composed of either of adenine (A), guanine (G), thymine (T), or cytosine (C) [4]. As the main purpose of sequence alignment algorithm, permanent change in the nucleotide sequence, denoted as mutations, should be located [2]. Mutations are classified either as Small-scale mutations or as Large-scale mutations [5]. A small scale mutation which affect a small gene in one or a few nucleotides causes a nucleotide base substitution, insertion, or deletion of the genetic material, DNA or RNA as shown in Fig. 1. The mutations, which happen randomly, are being recognized as particularly important in many genetic disorders. Therefore, detecting and locating point mutations may be known as the most important challenge for detecting genetic diseases [6].

The rapid growth of DNA sequence data in biology databases has necessitated efficient collating, organizing, identifying, retrieving, and searching the sequences as DNA representation. It's worse when a large amount of data should be processed in real-time [7]. However, considering big-data processing, traditional electronic computers suffer from many limitations including high power consumption, heat generation, high delay and slow response [8, 9].

Various sequence alignment methods have been proposed so far [10], which are categorized into two main groups; Global alignment algorithm, such as Needleman–Wunsch [11] which finds the most probable location of large number of short-length read sequences in a whole genome. On the other hand, BLAST [12] and Smith-Waterman [13] algorithms taking advantages of dynamic programming [14], are popular local alignment algorithms which locate the mutations by comparing each matched read sequence with its corresponding part in the reference genome. The main limitation of these algorithms is their computational time which increases by increasing the length of input DNA string [15]. Although proposing various speedup mechanisms [16, 17] could improve the computational speed, these algorithms still suffer from the limitations of electrical processing.

As another approach toward DNA sequence alignment, researchers have considered graphical representations of DNA sequences [18] to take advantages from image processing [19] capability to extract useful results from massive native data which aids in sequencing and annotating genomes. In graphical representations [20,21], using vectors for coding nucleotides, the DNA sequences are presented in the form of curves in two or more dimensional space; while the resemblance is specified by overlapping the two coded curves of DNA sequences. The main advantage of graphical representation methods is providing a visual inspection of data, helpful in recognizing, searching, organizing, and analyzing DNA sequences. But graphical representations are attended by various limitations. These limitations include huge computational complexity, representation complexity, difficulty of observation in case of multi-dimensional graphical representation [22-25], loss of information, degeneracy, difficulty of observing the coded curves, and difficulty of visualization of long DNA sequences. Specifically, the loss of information occurs when the resultant curve is crossed and overlapped by itself. The degeneracy is occurred when different nucleotide sequences are coded by same curves. Scrolling the coded curves needs a large and accurate processing. This effect is made worse when the DNA sequence becomes long.

**Fig. 1.** Point mutation types in DNA sequence alignment.

As a powerful 2D graphical method, DV-Curve (Dual-Vector Curve) coding method [26], by avoiding degeneracy and loss of information, avoids self-crossing and overlapping of coded DNA curves and offers good visualization to represent long sequences in 2D space. As key advantages, it reflects DNA sequence length and enables retrieving each DNA character from the coded curve. DV-curve representation specifies the resemblance by overlapping the two 2D DNA coded curves. Then the differences can be located visually. Analysis of the resultant image to declare the differences takes a long time and consumes high electrical power by sequential operation of electrical computers. In spite of its advantages, DV-curve method has not been get powered as a famous DNA alignment technique due to is low speed and incomprehensible output. This problem has been addressed in the work through optical implementation.

To overcome electrical computer limitations in implementing DV-curve method, optical computation has been proposed [27-29]. Capability of parallel processing in optical computers motivates us to optically implement a global and local sequence alignment method. For this purpose, optical correlation proposed in [30, 31] benefits from the potential of high speed processing can be employed to perform rapid global alignment for detecting high similarity among various sequences.

Moreover, a novel method, built upon DV-curve, is proposed as a local sequence alignment method to analyze genome sequences in five steps: DNA coding, alignment, noise cancellation, simplification, and modification, and benefits from wavelength, polarization and amplitude of the optical signals. In this method, traditional DV-curve method is extended to analyze optically coded DNA sequences based on novel coding method adopting amplitude, polarization, and wavelength of the signals. Once repeated in horizontal and vertical directions, two coded DNA curves are cross-matched to locate probable character mismatches in the alignment phase. The noise of output is eliminated by an optical thresholder in the noise cancellation phase. The output of the method is fed to a cylinder and degrader to produce an accurate and easy understanding output in the simplification and modulation phases. Despite the limitations of series computer to analyze the DV-curve output, the proposed architecture is capable of immediately detecting and locating any character deletion, insertion, or mutation in the input DNA sequence compared to the reference sequence.

In all, the main contributions of this paper are summarized as follows:
- Proposing a novel optical method for curve coding of DNA sequences that benefits from Dual-Vector, amplitude, polarization, and wavelength coding of the optical signal.
- Design an extended DV-Curve method to locate character matching, insertion, deletion, and mutations in comparison of two genome sequences.
- Developing a low-cost all-optical architecture for the proposed DV-Curve method to provide parallel DNA sequence alignment.
- Design a novel graphene based tunable color filter for optical processing.

The organization of the paper is as follows; an optical correlation-based global alignment structure and a proposed optical structure performing local alignment are presented in subsections 2.A and 2.B, respectively, while the DV-curve method is reviewed in subsection 2.B.1. Section 3 proposes a novel algorithm for local sequence alignment adopting the extended DV-Curve method; while its parallel optical structure is presented in Section 4. Section 5 presents simulation results of DNA sequence alignment in LUMERICAL frameworks. The implementation costs are discussed in Section 6. Finally, Section 7 concludes the paper and discusses the future work.

## 2. METHODOLOGY

The general structure of the proposed optical DNA alignment system, as shown in Fig. 2, includes two main parts: global alignment algorithm, built upon the method of optical correlator [30-32], and proposed local alignment algorithm, built upon the DV-curve method, and extended optically. Outputs of the proposed optical global alignment structure are fed to the proposed optical structure implementing local alignment procedure.

### A. Global Alignment Structure

Global alignment is obtained using a traditional optical correlator, as discussed in [30-32]. The correlation process is performed in Fourier domain [28]. At the first step, reference and query sequences are coded using an appropriate amplitude coding scheme. Then, the Fourier transform is computed for the reference and query sequences. Afterwards, the Fourier transform of the correlation between the reference and query sequence is calculated by multiplying their computed Fourier transforms. Finally, an inverse Fourier transform of the multiplication result is conducted that leads to correlation output. The resultant correlation output locates regions with the highest similarity corresponding to approximate query location in the reference genome with a peak correlation value [30]. In other words, global alignment module is responsible for finding these approximate locations. The approximate query locations are fed to the next step, as local alignment, to exactly locate any probable mutations.

### B. Local Alignment Structure

Once global alignment is performed, local alignment should be carried out to compare each read sequence with its corresponding matched part in the reference genome to detect and locate the mutation types, such as substitution, insertion and deletion. The local alignment is performed by an optical extended DV-curve method, named as Hybrid Amplitude-Wavelength-Polarization Optical DV-Curve (HAWPOD) method, to exactly locate every type of mutations.

As follows, the structure of traditional DV-curve method is reviewed thoroughly. Then HAWPOD method and its optical implementation are present as a novel integration of local DNA sequence alignment methods in details in Section 3 and 4.

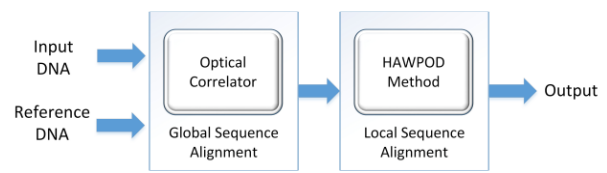

**Fig. 2**. Optical DNA correlation used for global DNA sequence alignment and proposed optical extension of DV-curve method for local DNA sequence alignment.

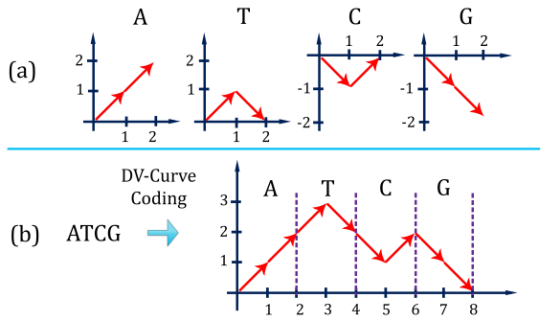

**Fig. 3**. Traditional DV-Curve coding scheme.

*B.1. DV-curve Method*

In this subsection, the structure of DV-curve [26] is explained. At first, each nucleotide (i.e. each of A/C/G/T character) is coded by two different vectors, as shown in Fig. 3. Specifically, every character extends two points along horizontal axes, therefore, the resultant DV-curve coded sequence reflects the length of DNA sequence as calculated in Equation 1. It also enables locating each character on DV-curve coded sequence. It should be noted that the character A extends two upward points, and character G extends two downward points; while, the extension for T is up and downward points, and extension for C is down and upward points. The corresponding DV-curve coded sequence is formed by connecting all the vectors one by one, as shown in Fig. 3.b. Since DV-curve coding uses two horizontal pixels for each character and connects them one by one in 2D space, every two pixels present one nucleotide. Therefore, DV-curve coding scheme enables locating each character of DNA sequences on the corresponding 2D curve. So, as a key advantage of this coding method, it can retrieve DNA characters from the coded curve. In this manner, as shown in Fig. 3.a, adopting DV-curve coding method, coded A/T/G/C characters never cross and overlap by itself, so, the DNA coded curve never has a circuit in its resultant 2D graph, and hence, avoids loss of information and degeneracy. The latter property empowers the DV-curve to visualize long DNA sequences, as shown in Fig. 4. Moreover, the resultant coded curve is not affected by the length of sequence. As an example, long sequences with 150 and 300 base pairs are visualized in Fig. 4.a and Fig. 4.b, respectively.

$$n = \frac{X_{end}}{2} \qquad (1)$$

The DV-curve method specifies the resemblance of two DNA sequences by overlapping their corresponding 2D curves. Two curves are overlapped exactly when two DNA sequences are identical. However, when a mutation is occurred, two DV-curve coded sequences become non overlapped and two curves take some distance from each other. In this manner, every mutation makes two coded DNA curves closer or further away from each other. So, variation of distance between two DNA curves expresses difference between the corresponding DNA sequences, as shown in Fig. 5. This figure depicts the impact of various types of mutation on DV-curve outputs. While character substitution only makes two coded curves closer or further in vertical direction, as shown as vertical movement problem in Fig. 5.a, the insertion and deletion types of mutation make them closer or further in horizontal and vertical directions, respectively, shown as horizontal movement problem in Fig. 5.b and Fig. 5.c. In this manner, utilizing DV-curve method, the differences between two DNA sequences can be detected and located immediately by exploring and observing the resultant curves, but declaration of the differences takes a long time and consumes large amount of electrical power.

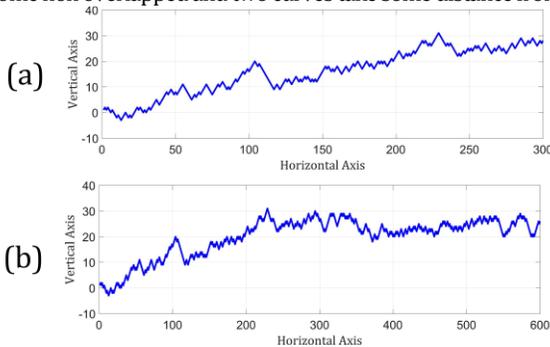

**Fig. 4**. DV-Curve of DNA sequence in length (a) 150 base-pair and (b) 300 base-pair.

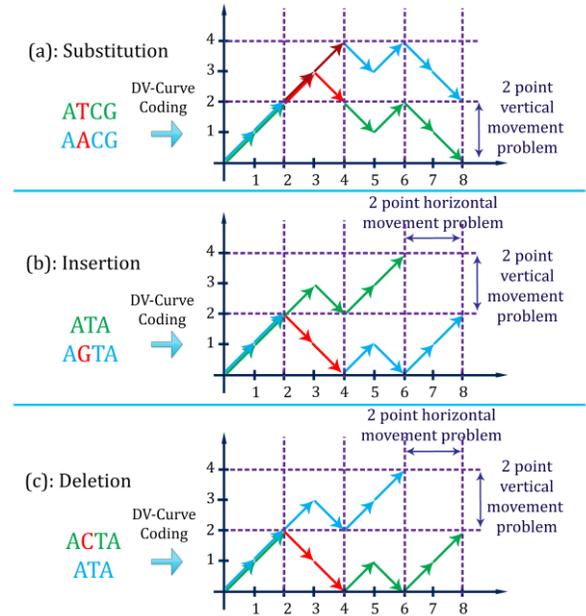

**Fig. 5**. Analysis of DV-curve output in case of single base-pair (a) substitution, (b) insertion, and (c) deletion.

## 3. PROPOSED HAWPOD METHOD

The HAWPOD structure utilized for DNA sequence alignment is composed of five main parts as shown in Fig. 6. As illustrated in this figure, as the first step of the local alignment phase, HAWPOD coding module encodes the input and reference DNA strings as 2D images adopting DV-curve coding method utilizing polarization, wavelength, and amplitude modulation of the optical signal. At the next step, the coded curves are repeated and overlapped on each other by color filter based on HAWPOD alignment method. Afterward, undesired noise at the primary output, due to sequence overlapping and non-ideal color filter, is eliminated by an optical thresholder in the noise cancellation phase. Then in the simplification phase, the output is converged in a row by an optical cylindrical lens to simplify the output and make the analysis easier. Finally, output of the HAWPOD method is modified by feeding to a grating with diffraction order of two to present an accurate output. Therefore, the HAWPOD technique can detect and locate any character mutation which is explained in as follow in details.

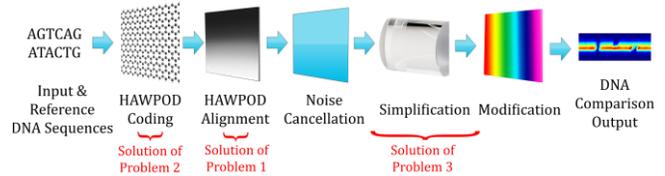

**Fig. 6**. Proposed optical local sequence alignment by HAWPOD method.

For representing an accurate and easy to understand output, the proposed HAWPOD structure should address following problems faced by traditional DV-curve methods:

- **Problem 1**: Any mutation in the input sequence prevents curve overlapping with the coded curve of reference genome afterwards.
- **Problem 2**: Undesired matching of nucleotides may happen due to character insertion, deletion, or substitution.
- **Problem 3**: Complex and noisy output image, as a result of sequence overlapping and non-ideal color filter result, necessitates further processing of DV-curve output.

As follows, we discuss the solutions HAWPOD method proposes to overcome the aforementioned problems.

### A. Solution of Problem 1: HAWPOD Alignment Method

**Problem:** As discussed before, for DNA sequence alignment by the traditional DV-curve method, two DV-Curves of DNA sequences are overlapped with each other, as shown in Fig 7.a. Two overlapped coded DNA sequences have intersections and result in bright pixels in output, as long as no mutation exists. Once a mutation occurs, as shown in Fig 7.b, output is corrupted starting from the place of mutation and bright pixels disappear, since the corresponding curves of DNA sequences get away from each other starting from the place of mutation, due to the different two-point horizontal movements for insertion and deletion, in addition two point vertical movements for various types of mutations. Analyzing the resultant output curve to clarify the differences needs a complex computation. To address the problem, we propose an optical approach to analyze the output of DV-curve.

**Solution:** For avoiding the loss of bright points while overlapping two coded curves, the coded reference DNA curve is repeated in vertical direction both upward and downward the main curve, as shown in Fig. 8.b, to address vertical movement problems in the case of various mutation types, as discussed above. Although vertical repetition enables exact locating of character substitutions, it cannot address the horizontal movement problem in the case of character insertions and deletions.

For example, as depicted in Fig. 8.a, 14th character of two DNA strings is considered for functional analysis of HAWPOD's output in the case of character substitution, insertion, and deletion. Fig. 8.c depicts successful locating of character substitution utilizing vertical repetition of reference curve in HAWPOD method; while it fails in the case of character insertion and deletion due to the problem of horizontal movement. To address this problem, the repeated coded curves are shifted in horizontal direction towards left and right of the main curve, as shown in Fig. 9. Therefore, character insertions and deletions, as well as character substitution, can be located, as shown in Fig. 8.d. But the aforementioned horizontal shift eliminates the two points horizontal gap in the case of character deletion, as shown in Fig. 8.d, while the gap is required to locate character deletion in the simplification phase, explained as follows.

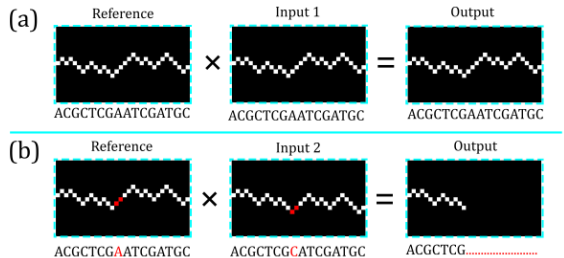

**Fig. 7**. Overlapping tradition DV-Curve of DNA sequences.

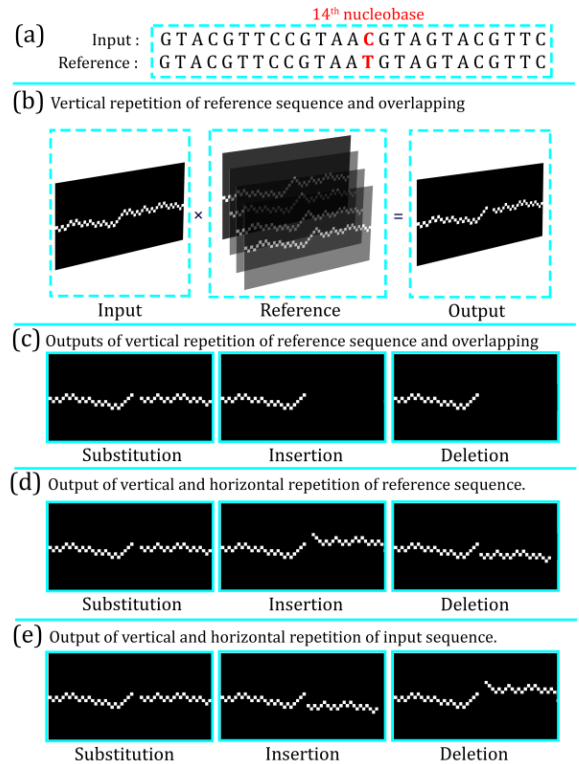

**Fig. 8**. HAWPOD output in all types of mutations.

To address the problem of two points horizontal gap in the case of character deletion, the whole structure is duplicated. As the first stage, the coded reference curve is repeated vertically and horizontally as discussed above; while at the second stage, the coded input curve, rather than the reference curve, is repeated vertically, as well as horizontally, as shown in Fig. 9. So, two points horizontal gap exists in the case of character deletion, as shown in Fig. 8.e. In this case, the substitution and deletion are determined while two-point horizontal gap in insertion is eliminated. Therefore, character substitutions and insertions are located by repeating the coded reference curve at the first stage; while the character substitutions and deletions are determined by repeating the coded input curve at the second stage. So, duplication of proposed method is capable of detecting and locating all mutation types required for local DNA sequence alignment. As the key advantage of the proposed method, two aforementioned processes are done in parallel, saving computation time.

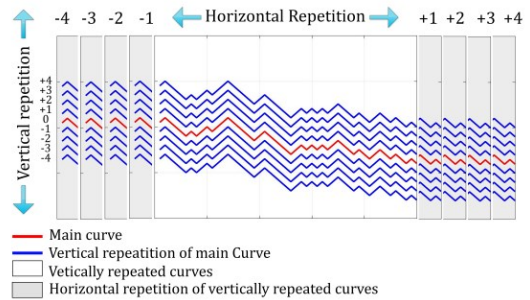

**Fig. 9**. Horizontal repetition of vertically repeated curves.

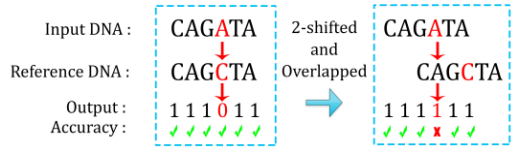

**Fig. 10**. Undesired non zero-pixel in output by repetition of curve of DNA sequence.

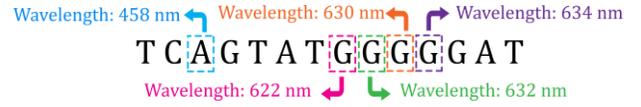

**Fig. 12**. Example of HAWPOD coding scheme.

### B. Solution of Problem 2: HAWPOD Coding Scheme

**Problem:** The traditional DV-Curve takes advantages of dual-vector coding of four possible characters, as shown in Fig. 3. It's worth noting that although each code set, defined in this figure, composed of different vectors with the same value, repetition of coded DNA sequences may result in undesired non-zero output in the case of string mismatch caused by horizontal movement of coded DNA sequences. Fig. 10 depicts the undesired non-zero points in resultant output of two DNA overlapping, as a result of two horizontal shift of coded input DNA sequence in the case of string substitution.

**Solution:** To avoid undesired non-zero points in the resultant output string, we propose HAWPOD coding which exploit amplitude, polarization, and wavelength features of the optical signals. In this manner, each nucleotide, in additional to a distinct dual-vector coding inherited from traditional DV-curve by amplitude, is coded on a distinct polarization and wavelength. The polarization angle varies in the range of [0, 180] which results in modulation wavelength in the range of [450nm, 650nm]. The wavelength is defined by the nucleotide itself, as well as its pre nucleotide and next nucleotide. Equation 2 presents our HAWPOD coding scheme for representing A, T, G and C nucleotides in a DNA sequence. As a key advantage of the proposed coding scheme, wavelength of each nucleotide dynamically changes for contiguous repeated nucleotides. The HAWPOD coding method adopts modulation wavelength for a nucleotide ($W_{Ni}$) as follows:

$$W_{Ni} = w_{Ni} - (k_{(Ni.Ni-1)} \times 8) - (k_{(Ni.Ni+1)} \times 2) + ((R-1) \times 2) \quad (2)$$

where, $N_i$ is the current nucleotide in position $i$; $N_{i-1}$ is its previous nucleotide, and $N_{i+1}$ is its next one. Without loss of generality, the value of $w_{Ni}$, as the modulation wavelength of various characters, is set to 480 $nm$ for A, 530 $nm$ for T, 580$nm$ for C, and 630$nm$ for G nucleotides to provide a non-overlapping coding of various nucleotides. Values of parameter $k_{(m.n)}$ for various values of *(m, n)* are provided in Table 1.

**Table 1: k values for various m,n**

| m\n | A | T | C | G |
|---|---|---|---|---|
| A | 0 | 1 | 2 | 3 |
| T | 1 | 0 | 3 | 2 |
| C | 2 | 3 | 0 | 1 |
| G | 3 | 2 | 1 | 0 |

Finally, *R* is the number of contiguous repeated nucleotides in a DNA sequence; it equals 1 for the first unrepeated presence of each nucleotide $(i)$. If the corresponding nucleotide is repeated in the next position$(i' = i + 1)$, with same previous$(i' - 1 = i - 1)$ and next$(i' + 1 = i + 1)$ nucleobases, the value of *R* is increased by 1. As obvious, *R* increases if the same nucleotides are chained continuously in a DNA sequence like "AAAAAA", where each character A and its pervious character have the same previous and next nucleotides. Maximum value of *R* is assumed to be 10, while it is initially set to 1.

According to Equation 2, modulation wavelength of various nucleotides varies in the range of [450$nm$ − 650$nm$]; while consequent wavelength channels are $2nm$ apart in average due to the frequency dependent free spectral range. As shown in Fig. 11, the proposed wavelength assignment approach provides 100 distinct codes adopting [0 − 180] different polarization angles which is required for coding all possible nucleotides. This color filtering is achieved through changing polarization angel which will be discussed in Section 4.

For example, as shown in Fig. 12, the proposed vector coding method, empowered by polarization coding, encodes nucleotide A, with C as its previous nucleotide and G as its next nucleotide with modulation wavelength of 458$nm$ based on Equation 2 with $k_{A.C} = 2. k_{A.G} = 3. and\ R = 1$. On the other hand, modulation wavelength of first occurrence of nucleotide G with nucleotides T and G as its previous and next nucleotides, respectively, is 622$nm$. For the first repetition of nucleotide, G, with G as its previous and next nucleotides, we have $R = 1$. Hence, according to Equation 2, modulation wavelength is chosen as 630$nm$; while for the second and third contiguous repetitions of G, with the same previous and next nucleotides it is chosen as 632$nm$ and 634 $nm$.

Moreover, extending DV-Curve coding with different modulation wavelengths and polarization, as proposed in HAWPOD coding method, we can prevent the undesired non zero output at the output image.

Based on HAWPOD coding scheme, the references and input DNA sequences are coded and represented as a chain of vectors with specific dual-vector coding, amplitude, wavelength, and polarization. At the next step, the coded DNA curve, which is repeated horizontally and vertically, is overlapped with another coded DNA curve. So, a bright point appears in the resultant output in the case that two crossed points are identical, which imposes the same amplitude, wavelength, and polarization of the corresponding 2D vectors. It should be noted that utilizing different modulation wavelengths and polarization for optical coding eliminates data interference in horizontal repetitions.

### C. Solution of Problem 3: Output noise cancellation, simplification, and modification

**Problem:** As depicted in Fig. 8.d and Fig. 8.e, the primary outputs of HAWPOD method are presented as 2D curves, while scrolling the curves in 2D space needs an accurate process to analyze and clarify the exact location of mutations. Moreover, crosstalk resulted from non-ideal color filter would affect the performance of the system in the optical implementation.

**Solution:** For achieving an easy to understand output, the primary output, achieved through HAWPOD coding and aligning procedures, is first optically thresholded to eliminate the crosstalk noise resulted from non-ideal color filter in the primary output, as shown in Fig 13.b. It should be noted that the crosstalk noise affects performance of the system. Then, a cylindrical lens is used to present the output as a single line, as depicted in Fig. 13.c, to avoid complex 2D curves The optical thresholding is necessary to avoid undesirable noise prior to signal feeding to the cylindrical lens.

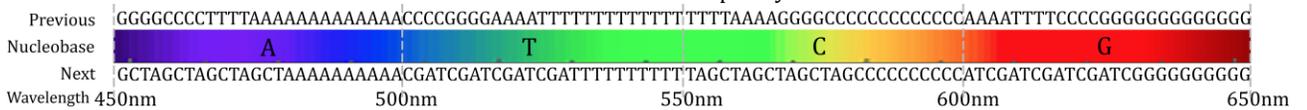

**Fig. 11**. HAWPOD scheme for polarization and wavelength coding of nucleotides A, T, C, G.

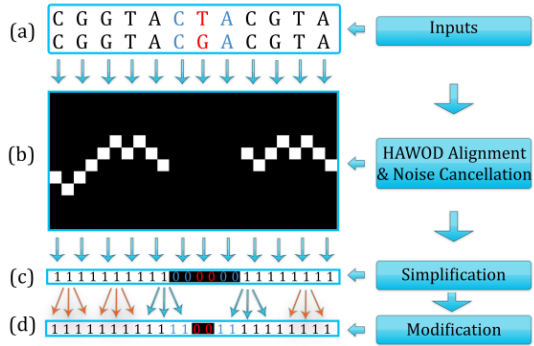

**Fig. 13**. Step-by-step outputs in HAWPOD method.

Based on HAWPOD coding scheme, coding of each nucleotide in a DNA strings is affected by its previous, as well as its next nucleotides. So, in the case of character mutation, coding of three consequent nucleotides would be altered. In this manner, the proposed wavelength assignment approach for modulating each nucleotide in HAWPOD coding scheme causes two undesired mismatches for pre and next nucleotides in outputs of all types of mutations, as shown in Fig. 13. For eliminating undesired zero points in the resultant output curve, referred to as the deficiency of HAWPOD coding scheme, we used a grating with diffraction order of two to map every bright pixel to five pixels in a row. It compensates the missed matching points of the previous and next nucleotides in the case of mutations, as shown in Fig. 13.c, while the final output of the grating with diffraction is shown in Fig. 13.d.

Converging the corresponding outputs of the two consequent stages of HAWPOD method, the final output is appeared in two lines. Where, the first and second lines are represented by repetition of reference and input DNA sequences in the first and second stages of the duplicated structure, respectively. It should be noted that the number of columns of the output matrix represents length of the coded DNA string as Equation 1. As shown in Fig. 14, binary values of '11' in both first and second rows of the output matrix represent character matching among input and reference DNA sequences. While, '00' values in both rows depict characters' mismatch, known as substitution. In the latter case, binary values of '00' in the first row beside binary values of '11' in the second row represent insertion of the corresponding characters in input DNA string. Finally, binary values of '11' in the first row beside binary values of '00' in the second row represent deletion of the corresponding characters. So, we can easily detect, as well as locate, character insertions, deletions, and mutation comparing two different DNA sequences.

In all, we can conclude that the proposed method can both detect and locate any possible character insertion, deletion, and substitution in the input genome string compared to the reference one. It is worth noting that the output of HAWPOD method is understandable and simple to analyze. Its accuracy, investigated through optical simulations in Section 5, confirms high precession of the HAWPOD method.

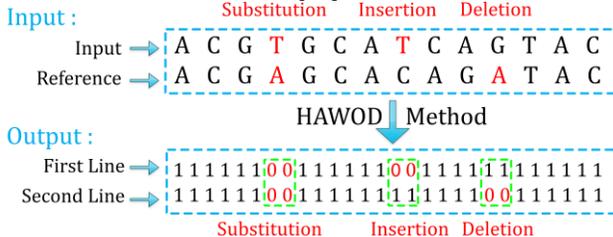

**Fig. 14**. Explanation of HAWPOD method output.

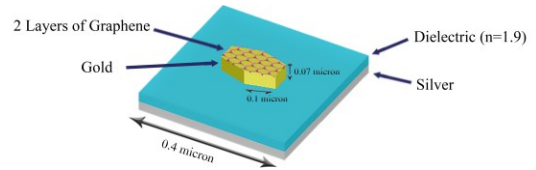

**Fig. 15**. Schematic of the proposed tunable graphene-based color filter cell consisting of two layers deposited on a gold nanostructure.

Advanced properties of HAWPOD are as follows:
- Simplicity and no need to set parameters.
- No circuit and degeneracy.
- No loss of information.
- Ability of retrieving every character from coded DNA through HAWPOD coding scheme.
- Indicating the length of DNA sequence by the horizontal length of output.
- Good visualization for long DNA sequence representation.
- Presenting an accurate and easy to understand output, eliminating further processing.
- Capable of detecting and locating any type of mutation.

## 4. OPTICAL IMPLEMENTATION OF HAWPOD METHOD

In the optical implementation phase of the proposed method, the references and input DNA sequences are coded with HAWPOD coding scheme. The HAWPOD coding method is implemented by an electrical tunable color filter and a polarizer for dual-vector, polarization, and wavelength coding part of HAWPOD. The color filter is composed of four layers, as depicted in Fig. 15. First layer is a metal layer which enables the structure working in the reflective mode. Then, a dielectric layer with refractive index of 1.53 is deposited on the metal layer, proceeded by an uniform graphene sheet placed on the top of the dielectric layer. Finally, the dielectric-graphene structure is repeated with a graphene sheet that looks like a fishnet.

We utilize a 2D graphene material model based on surface conductivity for graphene sheets. To work as a tunable structure, a voltage is applied to a polarization SLM located before graphene-based color filter. As the polarization of the incident field changes due to the applied voltage, the graphene based polarization spectrum is tuned in a way to work as a color filter [33]. The reflection spectrum of such color filter, with respect to the applied polarization, is shown in Fig. 16. This graphene-based optical modulator can provide 100 different band-pass filters with a wavelength dependent free spectral range of approximately 2 nanometers for each filter. These 100 band-pass filters are used as orthogonal codes in our HAWPOD coding scheme.

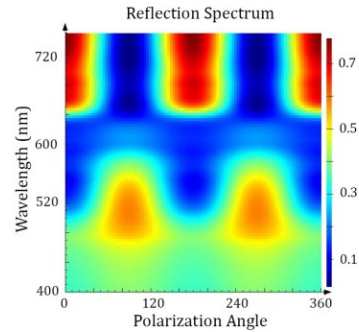

**Fig. 16**. Reflection spectrum of the proposed graphene-based color filter as the function of incident electric field polarization angle.

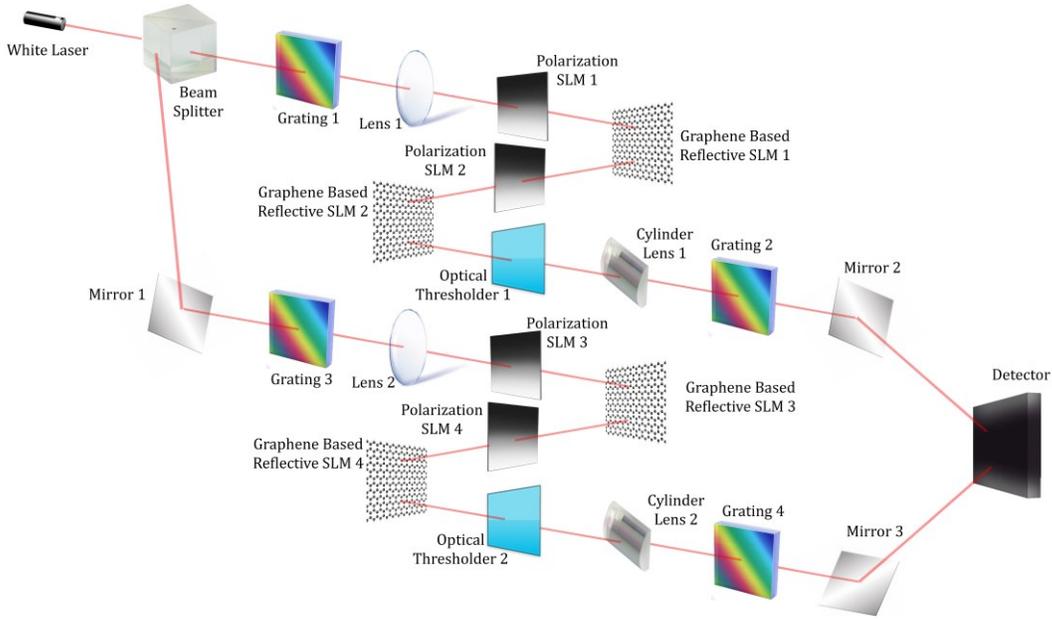

**Fig. 17.** Optical implementation of HAWPOD method.

The schematic of the proposed structure for DNA sequence alignment is depicted in Fig. 17. A white laser is illuminated on grating 1 and lens 1 for imagining purpose. To adopt the designed modulator in the proposed DNA alignment structure, the reference DNA sequence coded according to HAWPOD method is repeated in horizontal direction, as well as vertical direction, as the input signal.

Passing through the designed polarization SLM 1 (PSLM) as well as graphene-based reflective SLM 1 (GBRSLM), HAWPOD coded string of the input DNA sequence is further coded by color and polarization to represent the coded DNA sequence, respectively. Afterwards, the resultant optical signal, as the coded reference DNA sequence, is fed to the second color filter and polarizer (GBRSLM 2 and PSLM 2), which are coded based on the input DNA sequence. The corresponding optical signal passes through the filter and polarizer if the signals have the same wavelength as that of the coded input DNA sequences. Otherwise, the signals fail to pass and tend to disappear in the output.

Adopting the above structure, first line of the output image clarifies any possible character substitution and insertion. As discussed before, the proposed structure is duplicated in grating 3, lens 2, PSLM 3 and 4 and GBRSLM 3 and 4 to produce the second line of the output image to clarify substitutions and deletions, as well. Specifically, the second stage works as the first stage, while the input sequence, rather than the reference sequence, is repeated vertically and horizontally.

Finally, optical thresholding is performed by optical thresholders 1 and 2, shown in Fig. 17, to eliminate the crosstalk noises of the primary output. Optical thresholding prevents wavelength crosstalk responsible for the output noise originated from non-ideal color filter. To simplify the output image, cylindrical lenses 1 and 2 are used to coverage the output image in a row which make analysis and locating the mutations easier. To modify the output for addressing two undesired mismatching, a grating with diffraction order of two is used that maps every bright pixel to five pixels in row to overcome Problem 3 and provides an accurate output. Functionality and accuracy of the proposed optical structure implementing HAWPOD DNA sequence aligner is verified by numerical simulations in the next section

## 5. SIMULATION AND ANALYSIS

In this section, the correctness and accuracy of the proposed optical structure for global alignment are evaluated utilizing MATLAB simulation frameworks. At the next step, the correctness and accuracy of the proposed local alignment method, referred to as HAWPOD method, are verified in LUMERICAL FDTD simulation frameworks. For this purpose, at the first step, the proposed global alignment method compares input and reference DNA sequences against each other to detect the region of similarity with the highest score indicating as the correlation peak between two DNA sequences. Afterward, the corresponding output is fed to the HAWPOD system to detect and locate the mutations exactly. In the following simulation scenarios, the reference genome and queries are chosen from NA12878 human genome sequencing data. Both proposed global and local sequence alignment structures benefit from high speed processing offered by optical structures.

### A. Analysis of Global Alignment Procedure

In this section, the optical global alignment structure is evaluated through various simulations in MATLAB simulation framework. In this manner, a portion of PacBio Bioscience RS II technologies [34] is selected as the data-set for reference DNA sequence. For evaluating the result of correlation method adopted as the global alignment procedure, input DNA sequence is assumed as follow:

AGTTTGGCTCCTGTCAGCCTCCATAAAATCTGGGACGCCAAGAGCCCCACTGAGAGGTACAGGCTGGCCCTGTCTCGTAATGCATCTCGGTTAGCACAGGGGCTGATGTGACAGGCTGTAGGTTCCGTAACCCCTGCCATCTCAAGCATG

Correlation result of reference and input DNA sequences is depicted in Fig. 18. As shown in this figure, high similarity is detected at about 289th character of reference sequence. At the next step, a DNA subsequence, consisting of 150 characters starting from 289th character is extracted from the reference sequence for employing local alignment procedure, as will be discussed in the next section.

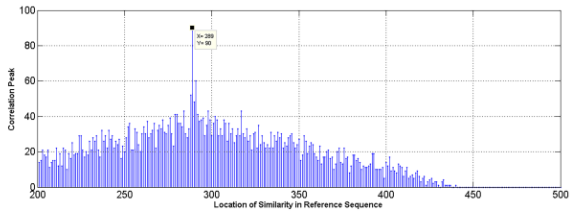

**Fig. 18.** Correlation result of reference and query sequences for global alignment

### B. Analysis of Local Alignment Procedure

In this section, the optical local alignment structure is evaluated. In this manner, two main simulation scenarios are considered. As the first scenario, single base pair mutations, such as single insertion, deletion, or substitution is considered because of the importance of single base pair detection required for disease diagnosing. In the second scenario, a multi base pair mutation is considered to confirm ability of HAWPOD method in detecting and locating various mutation types. Functionality and accuracy of the HAWPOD method and its optical structure are evaluated numerically with the commercially available software package LUMERICAL FDTD.

#### B.1. Analysis of HAWPOD method

For evaluation of HAWPOD method, two DNA sequences are compared against each other by the proposed optical global alignment method. Then, two 150-character length DNA subsequences, with highest similarity, are extracted by the global alignment procedure to be fed to HAWPOD structure for exact locating of various mutations. Two DNA sub-sequences, i.e. reference and input DNA subsequences, as depicted in Fig. 19.a, which include both single and multiple mutations, are as follow; including a single substitution, a single deletion, a single insertion, a multiple substitution, a multiple deletion, and a multiple insertion.

Input DNA subsequence =
AGTTTGGCTCCTG*G*CAGCCTCCATAAAATCTGGGACCCGAGCCCCAC
TGAGAGGTACAGGCTGG*A*CCCTGTCTCGTAATGCAGCTCGGTTAGCACA
GGGGC*AA*TGATGTGACAGGCTGTGGTTCCGTAACCTCCTG*TAT*TCTCAA
GCATG

Reference DNA subsequence =
AGTTTGGCTCCTG*T*CAGCCTCCATAAAATCTGGGACCC*AA*GAGCCCC
ACTGAGAGGTACAGGCTGGCCCTGTCTCGTAATGCAGCTCGGTTAGCAC
AGGGGCTGATGTGACAGGCTGT*A*GGTTCCGTAACCTCCTG*CCA*TCTCAA
GCATG

The corresponding DNA subsequences are optically coded according to HAWPOD coding method, proposed in subsection 3.B. For this purpose, each character is coded with in two horizontal pixels, so representing each DNA string requires 300 pixels in horizontal direction. Assuming uniform distribution of A/C/G/T characters through the DNA string, 120 pixels are required in vertical axis at the worst case. So a 2D input image of 120×300 pixels is produced by HAWPOD method for input and reference DNA subsequences.

To address Problem 1 and Problem 2, first and second stages of HAWPOD structure repeats corresponding 2D images of reference and input subsequences, respectively, in vertical and horizontal directions. Without loss of generality, in this case study, number of sequence repetitions in either direction of above, below, left, and right of the main curve is assumed to be 10. Then the resultant two 2D coded curves are cross-matched to produce primary HAWPOD output, presented in Fig. 19.b, which locates the mutation types. As illustrated in this figure, traversing the output image from left to right, the continuous line represents exact matching of two DNA subsequences, while each gap represents either single or multiple character mutations.

To address Problem 3, the undesired noise at the primary output is eliminated by a optical thresholder to present an accurate output as shown in Fig. 19.c. Then the output is fed to the optical cylinder lens to present a simple and more accurate result. Passing through a cylinder lens produces a 1×300 pixels image for each stage of HAWPOD method. So, two 1×300 pixel lines are produced, as shown in Fig. 19.d. To modify the output, the simplified output of HAWPOD is fed to a grating with diffraction order of two to map each bright pixel to five pixels in row in parallel. The resultant output is shown in Fig. 19.e, where the bright pixels verify continuous line detection at the output of HAWPOD method, which is interpreted as exact matching of two input strings. Moreover, dark pixels of the output image confirm existence of either character insertion, deletion, or substitution.

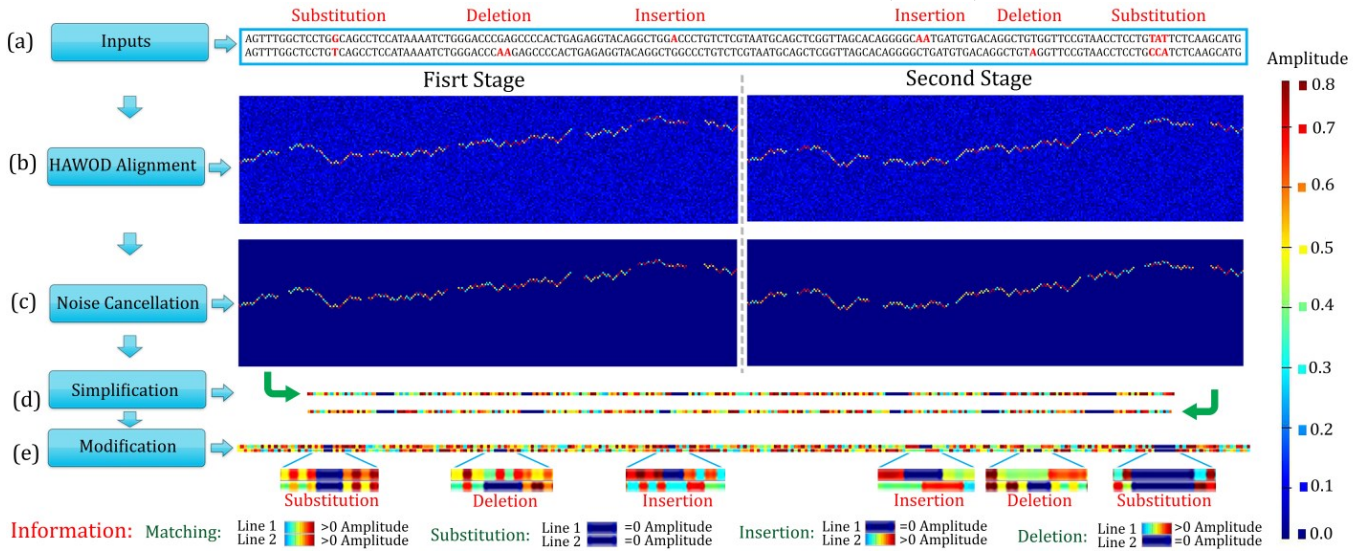

**Fig. 19** (a) Input DNA sequences, (b) Primary output of HAWPOD alignment phase, (c) Output of noise cancellation phase, (d) Output of simplification phase, and (e) final Output of HAWPOD method.

### B.2. Analysis of optical implementation of HAWPOD method

To validate accuracy of the proposed optical structure, we have implemented the designed wavelength and polarization filter in LUMERICAL FDTD as a numerical simulator. To this end, we model a color filter, composed of graphene sheets, gold, dielectric, and silver. The graphene sheet model is based on the surface conductivity of the graphene, while a conductivity scale of two is used for the graphene model to account for the two layers of graphene sheets used in our simulation [35]. To implement the dielectric, a simple model with constant refractive index of 1.9 is used, and the silver is assumed to be of perfect electric conductor boundary condition. Moreover, in our simulation, output of the SLM is modeled by calculating electric filed in front of the graphene based color filter at the corresponding wavelength. Wavelength of the proposed color filter output is determined by the applied polarization to the incident electric field [36].

Finally, we should analyze the crosstalk between neighboring pixels in the proposed SLM due to the electric field leakage of each pixel to its neighboring pixels and filters free spectral range. Fortunately, the crosstalk effect could be considered negligible, as can be tuned by adjusting the distance of graphene sheets, and does not affect our results due to the proposed coding scheme [37].

### B.3. Analysis of simulation output

Analyzing LUMERICAL FDTD simulation results, we can accurately locate single and multiple base pair insertions, deletions, and substitution in the reference sequence, compared to the input one. In HAWPOD method, the output is presented in two lines whose number of columns clarifies the location and type of mutations. Specifically, the zero amplitude of optical signal in both lines represents character substitution comparing two DNA sequences. Character insertion in the input sequence is presented by zero amplitude in the first line, while the amplitude of second line is greater than zero. On the other hand, character deletion is presented by zero amplitude in the second line, while the amplitude of first line is greater than zero. Finally, signal amplitude greater than zero in both lines presents sequence matching comparing two DNA sequences. It is also crystal clear from Fig. 19 that different wavelengths with a specific polarization have different reflectance.

For our case study, character insertions happen at 65th, 102nd, and 103rd nucleotides in input sequence, while its 39th, 40th, and 119th nucleotides are deleted, and character substitution are specified for 14th, 137th, 138th, and 139th nucleotides. The simulation result, depicted in Fig. 19.e, verify that both the HAWPOD method and the presented optical structure can locate character insertions, deletions, and mutation successfully, as expected.

For comparing output of HAWPOD method with that of traditional DV-Curve, we simulate the latter simulation in MATLAB simulation framework. To this end, the subsequences, used in LUMERICAL FDTD simulation, are fed to DV-Curve method in MATLAB. The resultant output, shown in Fig. 20, expresses that despite ability of good visualization in DV-Curve, the resultant output needs enormous graphical computation to clarify the mutations. So, it fails to present an accurate and easy understanding output.

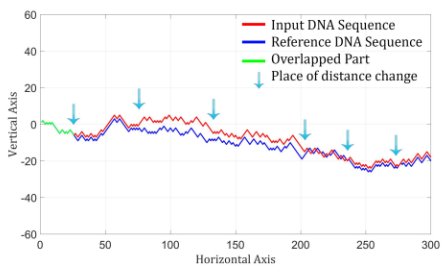

**Fig. 20**. Traditional DV-Curve output.

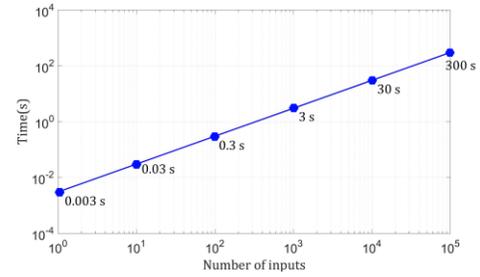

**Fig. 21**. Computational time of HAWPOD method for various number of inputs.

## 6. IMPLEMENTATION COST

After performing the global alignment, which takes 18 sec to align four million queries in whole genome sequence [32], local alignment should be performed to compare each read sequence with the reference genome to determine any possible mutations. Performing this comparison for a lot of reads suffers from high computational complexity and is time consuming. For example, adopting the well-known BLAST algorithm [12], this process requires estimated computational time of 12.28 seconds per read in average on a typical computer (i.e. an Intel® Core™ i7-4500U CPU @2.40 GHz with 8.00 GB of RAM) [38]. This process on HPC (i.e. Cisco® UCS Blade Server B200M2 × 2 Units with CPU 2 × 6 cores (2.4 GHz, 12 cores in total) and 96 GB of RAM) takes an estimated 0.53 seconds per read in average [38]. However, the DNA comparison in optical HAWPOD structure takes about 3 msec for each input, due to the switching time of 3 msec for a typical metamaterial-based SLM [39]. Summarizing the above discussion, Fig. 21 depicts the computation time of the proposed optical structure for various number of inputs. Finally, for processing 4 million inputs resulted from global alignment process, as discussed in [38], the BLAST algorithm takes running time of 568 days on a typical computer, and 24 days on an HPC. However, the proposed HAWPOD method only requires 200 minutes for comparing all 4 million inputs. Fig. 22 compares the computational time of HAWPOD method against two variant of BLAST methods for various numbers of inputs.

Regarding the memory requirements, the BLAST algorithm necessitates buffering large matrices, and so, not applicable for whole reads in DNA sequence alignment procedure. However, taking advantages of parallel processing nature of optics, the proposed optical structure eliminates the need of storage components, such as RAMs. So, the propose high-speed low-cost all-optical local DNA sequence aligner could perform the local DNA sequence alignment considerably faster taking advantages of parallel processing nature of optics.

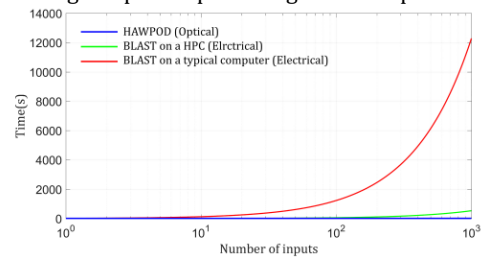

**Fig. 22**. Comparing the computational time for various number of inputs

# 7. CONCLUSION

In this paper, a novel optical structure for DNA sequence alignment has been proposed, named as HAWPOD. In this method, traditional DV-curve method is extended to analyze optically coded DNA sequences based on novel coding method adopting amplitude, polarization, and wavelength of the signals. Once repeated in horizontal and vertical directions, two coded DNA curves are cross-matched to extract probable character mismatches. The output of the method is fed to a cylinder and grating to produce an accurate and easy understanding output. The HAWPOD structure finds and locates character matches, as well as, character insertions, deletions, and substitutions comparing two genome sequences.

Parallel processing, inherently available by optical computing, motivated us to optically implement HAWPOD method. This allows us to benefit from both high speed and parallelism to locate probable mutations. Optical simulations, performed in LUMERICAL FDTD, verify correctness and accuracy of the proposed optical structure.

As a future work, physical realization of the optical structure implementing HAWPOD method would be addressed.